# A Comment on Nonextensive Statistical Mechanics

Oded Kafri

**Varicom Communications, Tel Aviv 68165 Israel.**

There is a conception that Boltzmann-Gibbs statistics cannot yield the long tail distribution. This is the justification for the intensive research of nonextensive entropies (i.e. Tsallis entropy and others). Here the error that caused this misconception is explained and it is shown that a long tail distribution exists in equilibrium thermodynamics for more than a century.

There are two typical distributions observed in the macroscopic world. The first one is the bell-like function, which is a result of the canonic distribution, and the second one is the long tail, which is a result of a power law distribution. While some statistical quantities are bell-like (the human height etc.), many others, like the human wealth etc. have a long tail distribution. The long tail distribution is as common in nature as the bell-like distribution.

Apparently, many believe that the long tail distribution cannot be obtained from equilibrium thermodynamics. The reason for this misconception is the way the canonic distribution is derived in some textbooks [1], namely to define a function as followed:

$$f(p_i) = \sum_{i=1}^{W} p_i \ln p_i + \alpha \sum_{i=1}^{W} p_i + \beta \sum_{i=1}^{W} p_i E_i \quad (1)$$

The derivation $\frac{\partial f(p_i)}{\partial p_i} = 0$ yields the canonic distribution. Here $p$ is the probability, $E$ is the energy, $\alpha$ and $\beta$ are Lagrange multipliers and $W$ is the number of microstates.

Eq.(1) looks exact, as the first term on the RHS represents the (-) Gibbs entropy, the second term is equivalent to the total number of particles, and the third term is the total amount of energy of the system. At a first glance, no approximations are made, and therefore, the only possible solution that maximizes the entropy for a given number of particles and a given amount of energy is the canonic distribution [1]. This implies that there is no way to obtain a power law distribution by maximizing Boltzmann-Gibbs entropy.

This is probably the reason for the enormous effort made to "generalize" the second law. The idea was to change the concept of entropy in a way that Eq.(1) will yield a power law distribution. This is the justification for Tsallis entropy [2], Renyi entropy [3], and more… The "entropy" of the highest impact is Tsallis entropy, which received, since it was suggested in 1988, according to Google scholar, more than 1250 citations. This warm welcome by the "community" is surprising as Tsallis entropy is nonextensive, which means a system in disequilibrium. The physical explanation for the nonextensivity is long-range interactions, which also implies disequilibrium. Therefore, accepting nonextensive entropy means giving up the most important concept of thermodynamics, namely the tendency of any system to reach equilibrium. In other words, nonextensivity means giving up the second law of thermodynamics altogether!

Hereafter, it is shown that the assumption that canonic distribution is the only solution that maximizes Boltzmann-Gibbs entropy under the constraints of Eq. (1) is erroneous.

Eq. (1) should be written,

$$f(p_i, p_j) = \sum_{j=1}^{W} p_j \ln p_j + \alpha \sum_{i=1}^{N} p_i + \beta \sum_{i=1}^{N} p_i E_i . \qquad (2)$$

Namely, Gibbs entropy should be summed over all possible different configurations $W$ of the ensemble (the microstates). However, the summation over the energies $p_i E_i$ should be done over the states $N$, as the distribution that we are looking for is the distribution of energy among states and not microstates (all the microstates have equal energy!). Usually, $W$ and $N$ are different numbers. An ensemble of $N$ states and $P$ particles where $P < N$, and no more than one particle is allowed in a state, has a number of configurations,

$$W(P, N) = \frac{N!}{(N-P)!P!} \qquad (3)$$

Applying Stirling formula and using Boltzmann entropy $S = \ln W$, we obtain that

$$S \cong -N\{p \ln p + (1-p)\ln(1-p)\}, \text{ where } p = \frac{P}{N}.$$

Or in Gibbs formalism,

$$S = -\sum_{j=1}^{W} p_j \ln p_j \cong -\sum_{i=1}^{N}\{p_i \ln p_i + (1-p_i)\ln(1-p_i)\} \qquad (4)$$

In the approximation $p \ll 1$, $(1-p)\ln(1-p)$ vanishes and the expression $-\sum_{i=1}^{N} p_i \ln p_i$ is entropy.

In this case Eq. (2) becomes, $f(p_i) \cong \sum_{i=1}^{N} p_i \ln p_i + \alpha \sum_{i=1}^{N} p_i + \beta \sum_{i=1}^{N} p_i E_i$ and yields the canonic distribution.

The conclusion up to this point is that the canonic distribution is not a law of nature, and it exists only at low occupation number systems.

Since, Eq. (1) is not always true, the legitimate way to look for other distributions is to calculate the number of microstates and their probabilities rather than changing the expression of the entropy.

Hereafter, it is shown that in fact, a power law distribution and its appropriate statistics exist in physics for over a century.

In the general case (neglecting degeneracy), we have to count all the configurations of $P$ particles in $N$ states for any value of $n$ (here we replace the symbol $p$ by $n$ as we allow $\frac{P}{N} > 1$). We follow the footsteps of Planck's seminal work from 1901 [4], namely,

$$W(P,N) = \frac{(N+P-1)!}{(N-1)!P!} \tag{5}$$

We apply again the Stirling formula as was done by Planck and obtain that $S \cong N\{(n+1)\ln(n+1) - n\ln n\}$, or in Gibbs formalism,

$$S = -\sum_{j=1}^{W} p_j \ln p_j \cong -\sum_{i=1}^{N}\{n_i \ln n_i - (n_i+1)\ln(n_i+1)\} \tag{6}$$

(Some may recall that this is Planck's derivation). If $n \ll 1$, we obtain again that the entropy is $-\sum_{i=1}^{N} p_i \ln p_i$, and therefore the canonic energy distribution is obtained as a private case. Since $n$ is now interpreted as a number and not a probability we omit the second term in Eq. (2). By substituting the entropy of Eq. (6) in Eq. (2)

$$f(n_i) \cong \sum_{i=1}^{N}\{n_i \ln n_i - (n_i+1)\ln(n_i+1)\} + \beta \sum_{i=1}^{N} n_i E_i \tag{7}$$

we obtain the Planck equation namely,

$$n_i = \frac{1}{e^{\beta E_i} - 1} \tag{8}$$

Similarly, substituting the entropy calculated from the number of microstates of Eq. (3) in Eq. (7) yields the Fermi-Dirac distribution.

We designate $\Phi_i = \beta E_i$ and we plot $\ln n_i$ versus $\ln \Phi_i$ and we see that when $n > 1$, Planck equation yields a power law distribution with a slope $-1$.

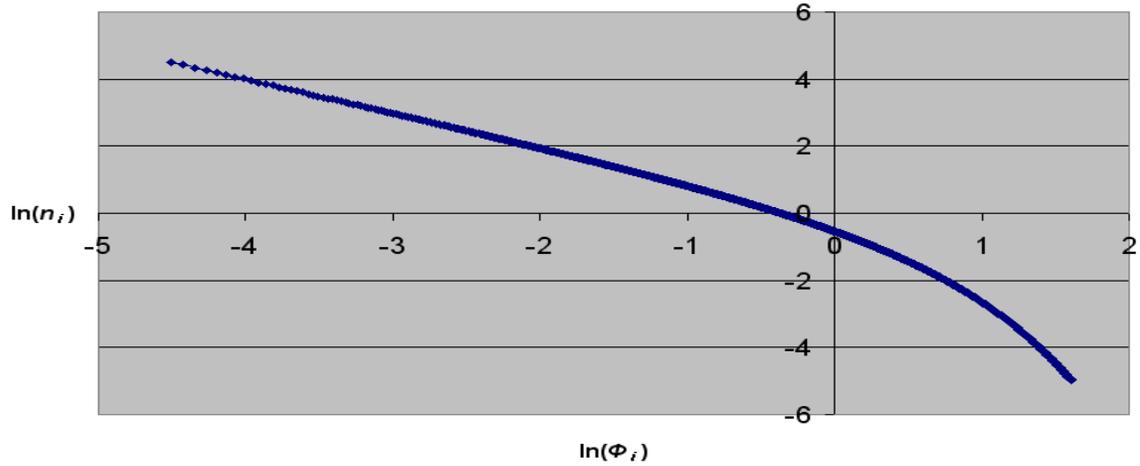

Fig.1 *A log-log plot of the occupation number versus the relative energy.*

In Fig.1 it is seen that when the number of particles is higher than the number of states (high occupation numbers), a power low distribution is obtained, and at low occupation numbers the canonic distribution is obtained. In the classic Rayleigh-Jeans approximation the distribution of photons in a radiation mode is a long-tail distribution. In fact, the same statistics was used recently to derive Benford's law and the wealth distribution [5,6,7].

References:

1. C. Back "*Generalize information and entropy in physics*" arxiv:0902.1235
2. C. Tsallis "*Possible generalization of Boltzmann-Gibbs entropy*" J. Stat. Phys. **52**,479 (1988).
3. E. K. Lenzi et. al. "*Statistical mechanics based on Renyi entropy*" PhisicaA **280** 337 (2000).
4. M. Planck "O*n the law of distribution of energy in the normal spectrum"* Annalen der Physik 4 553 (1901).
5. O. Kafri "*The second law as a cause for the evolution*" arxiv:0711.4507
6. O. Kafri "*Sociological inequality and the second law*" arxiv: 0805.3206
7. O. Kafri "*Entropy principle in direct derivation of Benford's law*" arxiv: 0901.3047